\pdftrailerid{}
\documentclass[sigconf]{acmart}

\copyrightyear{2026}
\acmYear{2026}
\setcopyright{cc}
\setcctype{by}

\usepackage{bm}

\newtheorem{assumption}{Assumption}
\newtheorem{definition}{Definition}
\newtheorem{proposition}{Proposition}
\newtheorem{theorem}{Theorem}
\newtheorem{remark}{Remark}
\newtheorem{corollary}{Corollary}

\newcommand{\R}{\mathbb{R}}
\newcommand{\E}{\mathbb{E}}
\newcommand{\dd}{\mathrm{d}}
\newcommand{\norm}[1]{\left\lVert #1\right\rVert}
\newcommand{\inner}[2]{\left\langle #1,#2\right\rangle}
\newcommand{\diag}{\operatorname{diag}}
\newcommand{\tapes}{\textsc{TAPES}}
\newcommand{\art}{\textsc{ART}}
\newcommand{\ays}{\textsc{AYS}}
\newcommand{\edm}{\textsc{EDM}}

\acmConference[Proceedings of the 7th ACM International Conference on AI in Finance (ICAIF 2026)]{7th ACM International Conference on AI in Finance}{November 14--17, 2026}{Milan, Italy}
\acmBooktitle{7th ACM International Conference on AI in Finance (ICAIF '26), November 14--17, 2026, Milan, Italy}
\acmDOI{10.1145/3858063.3858762}
\acmISBN{979-8-4007-3087-0/2026/11}
\AtBeginDocument{\pagestyle{standardpagestyle}\fancyhead[LE,RO]{}}
\makeatletter
\patchcmd{\maketitle}
  {{\itshape \acmConference@shortname, \acmConference@venue}\par}
  {\textbf{Publication.} \textit{Proceedings of the 7th ACM International Conference on AI in Finance}
    (ICAIF 2026), pp.\ 1135--1143. ACM, New York, NY, USA.
    Published 13 November 2026. Conference: Milan, Italy, 14--17 November 2026.\par}
  {}{\PackageError{publication-footnote}{ACM publication note hook not found}{}}
\makeatother
\begin{document}

\title{\tapes: Terminal Adjoint-Propagated Error Scheduling for Financial Diffusion Samplers}

\author{Yilie Huang}
\orcid{0000-0002-2922-9904}
\affiliation{%
  \department{Department of Applied Mathematics}
  \institution{The Hong Kong Polytechnic University}
  \city{Kowloon}
  \country{Hong Kong}
}
\email{yilie.huang@polyu.edu.hk}

\begin{abstract}
Financial diffusion samplers are often reused with a small number of function
evaluations (NFE) inside scenario and risk pipelines, where numerical time-grid
error can distort terminal quantities such as covariances, tails, and
drawdown proxies.  We study offline grid construction for a frozen probability-flow
ODE: the score network, conditioning, Euler solver, report normalization, and
deployment NFE remain fixed.  \tapes\ (Terminal Adjoint-Propagated Error
Scheduling) estimates signed local defects by Richardson comparisons,
transports them through the remaining discrete flow and a terminal-report
adjoint, and optimizes the squared norm of their sum before a capped direct
refinement.  We show that this transported signed sum is the first-order
terminal-report error and quantify, under an aligned-defect model, the
inefficiency of diagonal criteria.  Across six budgets on a regime-switching
simulation and chronologically held-out Mag7 contexts, \tapes\ gives the lowest
design- and audit-report errors in all 24 dataset--budget--report rows within
the reported five-pipeline panel.  Cross-budget mean relative audit-error
reductions are about 10--11\% on Regime-SV and about 1\% on Mag7, with
Bonferroni intervals excluding zero; all displayed
winners persist under a doubled-resolution reference, and a small Regime-SV
Heun probe preserves the \tapes-first audit ordering.  \tapes\ changes only the
offline schedule and retains the prescribed deployment NFE; claims concern
numerical fidelity to frozen generators.
\end{abstract}

\ccsdesc[500]{Computing methodologies~Modeling and simulation}
\ccsdesc[300]{Applied computing~Economics}
\keywords{financial diffusion models, low-NFE sampling, timestep scheduling,
goal-oriented error estimation, continuous-time mean--variance allocation}

\maketitle

\section{Introduction}
Diffusion models turn sampling into a sequence of denoising or
probability-flow steps \cite{ho2020denoising,song2021score}.  Financial
scenario generators include adversarial path models
\cite{wiese2020quantgan,liao2024sigwgan}, tabular diffusion
\cite{sattarov2023findiff}, and multivariate time-series diffusion
\cite{rasul2021timegrad,yuan2024diffusionts}.  Tail-aware scenario methods
target portfolio losses and tail measures
\cite{fairbrother2018scenario,cont2025tailgan}.  We study a frozen diffusion
sampler in such a scenario pipeline.  Two features
distinguish this setting from image generation.  First, the object consumed
downstream is not the raw path but a small set of terminal \emph{reports}:
covariances, volatility persistence, tail quantiles, drawdown proxies
\cite{cont2001empirical,bollerslev1986garch}.  Second, the sampler often runs
inside a larger Monte Carlo loop, so each scenario must be cheap
\cite{glasserman2003monte}; we study deployment budgets of 4--32 Euler
calls.  At such
budgets discretization error is material, and it is not spent evenly: two
local errors of similar Euclidean size at different denoising times can have
very different effects on a terminal report.  A grid that misallocates its
few steps silently biases every report the workflow consumes.

We therefore separate \emph{model error} from \emph{numerical schedule
error}.  The trained score network, optional conditioning, solver family,
report normalizers, paired noises, and deployment budget are fixed; only the
grid along the frozen probability-flow trajectory changes.  A high-NFE path
from the same generator is the paired numerical target---it is not asserted
to be the market law.  This yields a controlled estimand: how accurately can
$K$ Euler calls reproduce the terminal reports of the frozen generator?

\tapes\ treats grid construction as a goal-oriented numerical problem in the
dual-weighted-residual tradition \cite{becker2001optimal,giles2002adjoint}.
Adjoint-based ODE error analysis and timestep control already account for
propagated local errors \cite{utumi1996optimal,moon2003variational,cao2004posteriori};
later work also treats cancellation \cite{chaudhry2016posteriori}.  Related
goal-oriented local-error adaptation covers integral quantities of interest
\cite{meisrimel2020goal}.
A Richardson comparison estimates the signed defect created by each solver
interval \cite{hairer1993solving}; a backward adjoint transports the defects
through the remaining discrete flow and the terminal report; the objective
squares the \emph{sum} of transported defects, retaining cross-step
reinforcement and cancellation that additive local-error criteria discard.
A short, capped direct refinement of paired report error on development items
then refines each candidate, and one positive global grid per budget is
frozen for evaluation.  All probes, adjoints, and refinements are offline
construction costs; deployment uses exactly $K$ Euler calls.

\paragraph{Related work and positioning.}
Fast diffusion sampling includes hand-designed nonuniform clocks and
higher-order solvers \cite{karras2022edm,lu2022dpmsolver}.  Adaptive
Reparameterized Time (\art) casts timestep allocation as control of a
reparameterized sampling clock using a leading-order Euler-error objective
\cite{huang2026art,huang2026artneurips}.  Adaptive
time-stepping optimizes an error-bound objective over discretization points
\cite{chen2024adaptive}; AYS derives solver-, model-, and dataset-specific
schedules from stochastic-calculus objectives \cite{sabour2024align};
hierarchical schedule optimization combines initialization search with
midpoint refinement \cite{zhu2026hso}; DDSS and LD3 optimize sampler or grid
parameters through sample-quality or teacher-matching objectives
\cite{watson2022learning,tong2025ld3}.  Other grid criteria optimize solver approximation error
\cite{xue2024optimized}, score-derived transport work
\cite{williams2024scoreoptimal}, trajectory-based dynamic-programming costs
\cite{chen2025geometric}, or entropic time \cite{stancevic2025entropic};
instance-aware discretization conditions the grid on the input
\cite{yuan2026instanceaware}.  These are related construction
approaches, not evaluated rows in the common frozen-Euler financial panel.
The matched ablations below concern \tapes's own construction stages.  \tapes\
differs in what the objective measures: normalized \emph{terminal financial
reports}, reached through a transported-Richardson proposal that keeps signed
cross-step structure, followed by capped direct refinement.  At fixed solver
and NFE, we compare with \art, \ays, \edm, and UNIF under a common frozen
financial time-series sampler;
Section~\ref{sec:baselines} gives the exact implementations.

Continuous-time reinforcement learning (RL) connects stochastic control to
data-driven decision making \cite{wang2020rl,wang2020meanvariance}.
Financial applications include mean--variance portfolio selection
\cite{huang2022meanvariance,huang2024meanvariance} and asset--liability
management \cite{huang2025alm}.  Related linear--quadratic control studies
analyze exploration and regret \cite{huang2025lq,huang2026exploration}.  The allocator in
Section~\ref{sec:downstream} follows the mean--variance line.  \tapes\
constructs its grid by offline numerical optimization; RL enters only through
that downstream allocator.  Related offline control of a frozen diffusion backbone has also been
studied for image-inpainting guidance \cite{huang2026aid}.

\paragraph{Contributions.}
We make four contributions:
\begin{enumerate}
\item \textbf{Report-aware grid construction.} We optimize a positive grid by
transporting signed Richardson defects through a terminal-report adjoint,
squaring their sum, and applying capped direct refinement
(Section~\ref{sec:formulation}).
\item \textbf{Report-error theory.} We show that the transported signed sum
is the first-order term in terminal report error.  An aligned-defect model
then gives closed-form optimal grids and an inefficiency factor $\rho\ge1$
for adjoint-weighted diagonal criteria (Section~\ref{sec:analysis}).
\item \textbf{Frozen-sampler validation.} We test six budgets in two data
regimes, with joint uncertainty across budgets, doubled-resolution reference
checks, matched construction controls, and a transported-sum audit
(Section~\ref{sec:experiments}).
\item \textbf{Downstream sensitivity.} We trace a 21-day numerical bridge
and report an exploratory, post-selected allocation comparison under one
Mag7 protocol (Section~\ref{sec:downstream}).
\end{enumerate}

\section{Frozen Sampler and Report-Aware Grid Construction}
\label{sec:formulation}
\subsection{Frozen Sampler and Time-Changed Clock}
Let $Y=(y_1,\ldots,y_L)\in\R^{L\times m}$ be a future return window,
vectorized as $Y\in\R^d$, and let $\xi$ denote optional conditioning (lagged
returns, volatility summaries, identifiers).  \tapes\ applies to an
unconditional score model for windows $Y\sim p_{\rm data}$ or a conditional
model $s_\phi(x,\psi,\xi)\approx\nabla_x\log p_\psi(x\mid\xi)$; the Regime-SV
experiment uses the former, Mag7 the latter.  Either induces a frozen
probability-flow ODE
\begin{equation}
  \frac{\dd X_\psi}{\dd\psi}=F_\phi(X_\psi,\psi,\xi),\qquad \psi\in[0,T],
  \label{eq:pfode}
\end{equation}
with $\psi=0$ the noise endpoint and $\psi=T$ the generated window; \tapes\
never updates $\phi$.

We parameterize the schedule as a time change.  The \emph{diffusion clock}
$\psi$ is the trained flow's time; the \emph{computational clock}
$u\in[0,1]$ carries uniform solver effort with points $u_k=k/K$.  A monotone
$\tau:[0,1]\to[0,T]$ maps computation time to diffusion time; with
$Z_u=X_{\tau(u)}$,
\begin{equation}
  \frac{\dd Z_u}{\dd u}
  =
  \theta_K(u)F_\phi(Z_u,\tau(u),\xi),
  \qquad
  \frac{\dd\tau}{\dd u}=\theta_K(u)>0,
  \label{eq:twoclock}
\end{equation}
$\tau(0)=0$, $\tau(1)=T$.  A uniform grid in $u$ induces diffusion-time
intervals $\Delta_k=\int_{k/K}^{(k+1)/K}\theta_K(u)\,\dd u$: small
$\theta_K$ concentrates resolution, and positivity keeps the deployed grid
strictly increasing.

\begin{definition}[Admissible clock and deployed grid]
An admissible $K$-step clock is a positive integrable
$\theta_K(\cdot)$ with total mass $T$; its deployed grid is
$\psi_k=\tau(k/K)$.  Equivalently, a deployed schedule is a vector
$\Delta=(\Delta_0,\ldots,\Delta_{K-1})$ in the open simplex
$\mathcal D_K(T)=\{\Delta_k>0,\ \sum_k\Delta_k=T\}$, realized by the
piecewise-linear clock with speed $K\Delta_k$ on cell $k$.
\end{definition}

The implementation parameterizes the discrete clock by a floored softmax over
logits,
\begin{equation}
  \Delta_k
  =T\left[(1-\alpha_{\rm floor})
  \frac{\exp b_k}{\sum_j\exp b_j}
  +\frac{\alpha_{\rm floor}}{K}\right],
  \label{eq:softmax}
\end{equation}
which keeps the terminal-time constraint inside the parameterization; all
selected \tapes\ grids use $\alpha_{\rm floor}=10^{-4}$, and baseline
libraries retain their implemented floors.  We optimize one
context-independent clock per budget; development contexts only estimate
and select the objective.  When $K$ is fixed, we abbreviate $\theta_K$ by
$\theta$ below.

\subsection{Financial Reports and Paired Estimand}
Let $R_{\rm design}$ and $R_{\rm audit}$ be normalized terminal report maps.
The design vector contains per-asset means and variances, off-diagonal
covariances, lag-one autocovariances of squared returns, and smooth
peak-to-trough proxies of cumulative standardized returns (log-sum-exp
temperature 25); its dimension is 30 for the
five-asset Regime-SV window and 49 for the seven-asset Mag7 window.
Separately, the audit vector contains, per asset, a lag-one squared-return
autocorrelation, a leverage term, return autocorrelations at lags 1, 2, and 4,
skewness, kurtosis, 5\% return-tail quantile and below-quantile mean, and a
temperature-100 drawdown proxy; its
dimensions are 50 and 70, respectively.
These features are computed for each eight-observation window of training-standardized
returns.  The 5\% quantile uses linear interpolation; absent tied minima,
the below-quantile mean equals the window minimum.  Both are short-window
lower-tail proxies, with the return rather than loss sign convention.  Each
raw vector $\Phi$ is normalized as
$R(x,\xi)=\diag(\widetilde s_\Phi+10^{-6})^{-1}\{\Phi(x,\xi)-\mu_\Phi\}$,
where $\widetilde s_{\Phi,j}=s_{\Phi,j}$ if $s_{\Phi,j}\ge10^{-6}$ and 1
otherwise; feature moments are fitted on training windows and frozen.

Given a low-NFE grid $\Delta$ and a high-NFE reference generated from the
same initial noise and context,
\begin{align}
  E_{\rm design}(\Delta)
  &=\E\norm{R_{\rm design}(x_K^\Delta,\xi)-R_{\rm design}(x^{\rm ref},\xi)}_2^2,
  \label{eq:edesign}\\
  E_{\rm audit}(\Delta)
  &=\E\norm{R_{\rm audit}(x_K^\Delta,\xi)-R_{\rm audit}(x^{\rm ref},\xi)}_2^2 .
  \label{eq:eaudit}
\end{align}
All methods share context, initial noise, and reference path within each
paired item.  The design map supplies the terminal adjoint; the audit-family
map is also used by fixed-grid tuning, \ays, and the final
refinement/selection stage of \tapes\ on \emph{development} items.
Evaluation items are strictly held out (fresh noise for Regime-SV,
chronologically later contexts for Mag7), but the audit coordinate family is
not feature-held-out for every method; ``audit'' labels the report family,
not an untouched diagnostic.  The direct development objective for a
report map $R$ would minimize
$\E_{\rm dev}\|R(x_K^\Delta,\xi)-R(x^{\rm ref},\xi)\|_2^2$ over
$\mathcal D_K(T)$; \tapes\ approaches it with a perturbation-based proposal
followed by a small, capped number of direct-refinement iterations.

\subsection{Discrete Terminal-Adjoint Schedule Objective}
For a fixed deployed grid $\Delta$, write $x_k=x_k^\Delta$ and
$x_{k+1}=G_k(x_k)$, with $G_k(x)=x+\Delta_kF_\phi(x,\psi_k,\xi)$ for Euler.
With $\varphi_k^{\rm ex}$ the exact flow over interval $k$, the
signed local defect is
\begin{equation}
  e_k=\varphi_k^{\rm ex}(x_k)-G_k(x_k).
  \label{eq:defect}
\end{equation}
For $R:\R^d\to\R^p$ let $B_k=D_xG_k(x_k)$,
$A_K=J_xR(x_K,\xi)$, and $A_k=A_{k+1}B_k$; then $A_{k+1}e_k$ is the
first-order contribution of $e_k$ to the terminal report.  The
implementation evaluates $A_K$ by central differences (step $10^{-5}$) and
propagates through vector--Jacobian products of the full frozen Euler maps,
including the score-based drift \cite{griewank2008evaluating}.  \tapes\
estimates $e_k$ by Richardson comparison, sets
$w_k=A_{k+1}\widehat e_k$, and minimizes
\begin{equation}
  \mathcal L_{\tapes}
  =\E\left\|\sum_{k=0}^{K-1}w_k\right\|_\Lambda^2
  +\beta_{\rm diag}\,\E\sum_k\norm{w_k}_\Lambda^2
  +\rho_{\rm euc}\,\E\sum_k\norm{\widehat e_k}_2^2,
  \label{eq:tapesloss}
\end{equation}
with $\Lambda=I$ after normalization.  The leading term keeps signed
cross-time structure; the small diagonal and Euclidean stabilizers guard
against cancellation specific to a finite panel (either weight may be
zero).

\subsection{Implemented Construction and Baselines}
\label{sec:baselines}
\art\ \cite{huang2026art} is implemented here over the same positive-clock
family by minimizing the rescaled
Euclidean Richardson-defect objective
$\E_{\rm dev}\sum_k\|\widehat e_k^{(n)}\|_2^2
\{\Delta_k/(\Delta_k^{(n)}+10^{-12})\}^{2(r+1)}$
at outer iterate $n$ ($r=1$ for Euler): no adjoint, no cross-time term, no
report refinement, and selection by the defect objective itself (eight
candidates per $K$ for Regime-SV, 18 for Mag7).  UNIF is the exact uniform
clock.  \ays\ evaluates UNIF, a development-tuned one-parameter VP grid, and
a development-tuned Karras VP grid, tuning
$\lambda_{\max}\in\{1,2,3,4,6\}$ (and $\rho_{\rm Kar}\in\{3,5,7\}$ for the Karras
family) on 32 audit-family development items, best-of-three selected on four
of those items, each grid fitted once into the shared clock representation
with the optimizer cap set to zero.  \edm\ is the selected Karras VP grid.
All methods use the common frozen Euler sampler; Table~\ref{tab:finance-errors}
compares their end-to-end implementations at common deployment NFE, while
development data and offline compute follow each stated procedure.  The \ays\
row is the output of its development-only selection rule, and \edm\ reports
its Karras-family candidate separately.

\subsection{Two-Stage \tapes\ Optimizer}
At outer iterate $n$, a rollout gives $\widehat e_k^{(n)}$ and
$w_k^{(n)}$.  The proposal stage freezes these coefficients under the
local-order model
\begin{equation}
  \widehat e_k(\Delta_k)
  \approx
  \gamma_k^{(n)}\,\widehat e_k^{(n)},\qquad
  \gamma_k^{(n)}=\Big(\frac{\Delta_k}{\Delta_k^{(n)}+10^{-12}}\Big)^{r+1},
  \label{eq:coeffmodel}
\end{equation}
and minimizes the induced quadratic surrogate of \eqref{eq:tapesloss} over
$\mathcal D_K(T)$, safeguarded by the floor, damping, and a move limit.  The
evaluated pipeline uses 64 development paths; starts from four fitted base
clocks---one obtained from the \art\ objective, plus UNIF, the one-parameter
VP, and \edm---; takes one
proposal update per candidate; applies
a short L-BFGS-B refinement of direct paired audit-family error (capped at
eight iterations for Regime-SV, four for Mag7); and selects the candidate
with the lowest development error.  Candidate libraries contain three
(Regime-SV) or two (Mag7) tuples varying the cosine basis
size (6 or 8), floor ($10^{-4}$ or $10^{-3}$), stabilizer weights
($\beta_{\rm diag}\le10^{-2}$, $\rho_{\rm euc}\le10^{-3}$), damping, and
move limit, with proposal caps of 35--50 inner iterations; the selected
tuple varies with $K$.  Construction is offline; deployment uses $K$ model
calls.  Recorded per-budget construction wall times span 3.2--13.8 minutes
(Regime-SV) and 1.9--7.0 minutes (Mag7) on a laptop-class accelerator,
including all candidate grids and baseline tuning; this one-time cost can
be amortized when a frozen grid is reused across many generation calls.  The method is
relevant when low-NFE discretization error is appreciable and a high-NFE
numerical target is affordable; it cannot repair generator
misspecification.

\section{Perturbation Theory for \tapes}
\label{sec:analysis}
The analysis makes the clock reparameterization explicit and shows why
terminal report error is not an additive local-error density.

\begin{proposition}[Integrated-grid Euler defect]
\label{prop:defect}
Let $0=\psi_0<\cdots<\psi_K=T$ be a positive deployed grid and let $F_\phi$
be $C^2$ with bounded derivatives near the step.  With
$Q_\phi=\partial_\psi F_\phi+\nabla_xF_\phi\,F_\phi$,
\begin{equation}
  \varphi_k^{\rm ex}(x)-G_k(x)
  =
  \frac{\Delta_k^2}{2}Q_\phi(x,\psi_k,\xi)
  +O(\Delta_k^3),
  \label{eq:integratedclockdefect}
\end{equation}
and if the grid is induced by a $C^3$ time map with speed
$0<\theta_{\min}\le\theta\le\theta_{\max}$, then with $h=1/K$,
$\varphi_k^{\rm ex}(x)-G_k(x)
=\tfrac{h^2}{2}\theta(u_k)^2Q_\phi(x,\tau(u_k),\xi)+O(h^3)$.
\end{proposition}
\begin{proof}
Expanding the exact flow in diffusion time from $(x,\psi_k)$,
\[
\varphi_k^{\rm ex}(x)
=x+\Delta_kF_\phi(x,\psi_k,\xi)
+\frac{\Delta_k^2}{2}\frac{\dd}{\dd\psi}F_\phi\big(X_\psi,\psi,\xi\big)
\Big|_{\psi_k}+O(\Delta_k^3),
\]
and the total derivative along the flow is exactly
$Q_\phi=\partial_\psi F_\phi+\nabla_xF_\phi\,F_\phi$; subtracting
$G_k(x)=x+\Delta_kF_\phi(x,\psi_k,\xi)$ gives
\eqref{eq:integratedclockdefect}.  For the $h$-form, a $C^3$ time map has
$\Delta_k=\tau(u_k+h)-\tau(u_k)=h\theta(u_k)+\tfrac{h^2}{2}\dot\theta(u_k)
+O(h^3)$, so $\Delta_k^2=h^2\theta(u_k)^2+O(h^3)$; substituting into
\eqref{eq:integratedclockdefect} and using boundedness of $Q_\phi$ absorbs
the difference into $O(h^3)$.  Equivalently, in the $u$-clock the exact
reparameterized trajectory satisfies
$Z_{u_k+h}=Z_{u_k}+h\theta(u_k)F_\phi
+\tfrac{h^2}{2}\{\dot\theta(u_k)F_\phi+\theta(u_k)^2Q_\phi\}+O(h^3)$,
while the deployed integrated increment expands as
$\Delta_kF_\phi=h\theta(u_k)F_\phi+\tfrac{h^2}{2}\dot\theta(u_k)F_\phi
+O(h^3)$: the clock-acceleration terms $\tfrac{h^2}{2}\dot\theta F_\phi$
coincide and cancel in the defect, leaving only the
$\tfrac{h^2}{2}\theta^2Q_\phi$ term.
\end{proof}

\begin{remark}
\label{rem:naive}
If one instead applies left-endpoint Euler directly to the reparameterized
$u$-ODE with the approximation $\Delta_k\approx h\theta(u_k)$, the increment
omits $\tfrac{h^2}{2}\dot\theta(u_k)F_\phi$ and the local defect acquires
that additional clock-acceleration term.  \tapes\ deploys the integrated-grid step, so its defect is governed by
$Q_\phi$ alone.
\end{remark}

\begin{proposition}[Richardson sign and order]
\label{prop:richardson}
Let $r\ge1$ and suppose a smooth one-step method has signed defect
$d=c\Delta^{r+1}+O(\Delta^{r+2})$ under \eqref{eq:defect}.  With
$x_{k+1}^{\rm one}$ one full step and
$x_{k+1}^{\rm two}$ two half steps from the same state,
\begin{equation}
  \widehat e_k
  =\frac{x_{k+1}^{\rm two}-x_{k+1}^{\rm one}}{1-2^{-r}}
  =e_k+O(\Delta_k^{r+2}).
  \label{eq:richardson}
\end{equation}
\end{proposition}
\begin{proof}
The full step has defect $e_k=c_k\Delta_k^{r+1}+O(\Delta_k^{r+2})$ with
$c_k=c(\psi_k,x_k)$.  For the two half steps, the first contributes
$c_k(\Delta_k/2)^{r+1}+O(\Delta_k^{r+2})$.  The second starts from a state
$O(\Delta_k)$-close to $x_k$ at time $\psi_k+\Delta_k/2$, so by the $C^1$
dependence of $c$ its coefficient is $c_k+O(\Delta_k)$ and it contributes
$c_k(\Delta_k/2)^{r+1}+O(\Delta_k^{r+2})$; transporting the first-half
defect through the second half step multiplies it by
$I+O(\Delta_k)$, a further $O(\Delta_k^{r+2})$ change.  Hence the two-step
endpoint carries total defect
$2\cdot2^{-(r+1)}c_k\Delta_k^{r+1}+O(\Delta_k^{r+2})
=2^{-r}e_k+O(\Delta_k^{r+2})$.  Both endpoints approximate the same exact
endpoint $\varphi_k^{\rm ex}(x_k)$, so
$x_{k+1}^{\rm two}-x_{k+1}^{\rm one}
=\{\varphi_k^{\rm ex}(x_k)-x_{k+1}^{\rm one}\}
-\{\varphi_k^{\rm ex}(x_k)-x_{k+1}^{\rm two}\}
=(1-2^{-r})e_k+O(\Delta_k^{r+2})$; dividing by $1-2^{-r}$ gives
\eqref{eq:richardson}.
\end{proof}

\begin{assumption}[Regularity tube]
\label{ass:regularity}
Fix $K$.  On a tube containing the relevant trajectories, the step maps
$G_k$, exact local flows $\varphi_k^{\rm ex}$, and the differentiable design
report $R$ are $C^2$ with derivatives and transition products uniformly
bounded over the paired-item sampling law.  Let
$\epsilon_k=\|\varphi_k^{\rm ex}-G_k\|_{C^1}$ on the tube and
$\epsilon_{\rm def}=\sum_k\epsilon_k$, small enough that perturbations stay
in the tube; smooth Euler steps give $\epsilon_k=O(\Delta_k^2)$.  Constants
may depend on $K$.  The implemented drift clips score coordinates and clamps
the model-time argument at endpoint cutoffs, so it is piecewise smooth; the
assumption is local to branches away from these kinks, and the empirical
comparisons below do not rely on it.
\end{assumption}

\begin{theorem}[Terminal report error expansion]
\label{thm:adjointsum}
Let $x_K^{\rm ex}$ be the endpoint obtained by replacing each deployed step
by the exact flow over the same interval from the same initial state.  Under
Assumption~\ref{ass:regularity},
\begin{equation}
  R(x_K^{\rm ex},\xi)-R(x_K,\xi)
  =\sum_{k=0}^{K-1}A_{k+1}e_k+\mathcal R,\qquad
  \norm{\mathcal R}\le C_K\epsilon_{\rm def}^2 .
  \label{eq:adjointsum}
\end{equation}
\end{theorem}
\begin{proof}
Let $\delta_k$ denote the time-$\psi_k$ gap between the exact-flow
trajectory and the numerical trajectory, with $\delta_0=0$.  One step gives
\[
\begin{aligned}
\delta_{k+1}
&=\varphi_k^{\rm ex}(x_k+\delta_k)-G_k(x_k)\\
&=\varphi_k^{\rm ex}(x_k)-G_k(x_k)
+D\varphi_k^{\rm ex}(x_k)\,\delta_k+O(\|\delta_k\|^2)\\
&=e_k+B_k\delta_k+O(\epsilon_k\|\delta_k\|+\|\delta_k\|^2),
\end{aligned}
\]
using $\|D\varphi_k^{\rm ex}-DG_k\|\le\epsilon_k$ on the tube.  Because the
transition products $\|B_{j}\cdots B_{i}\|$ are uniformly bounded, unrolling
yields $\|\delta_k\|\le C\sum_{j<k}\epsilon_j\le C\epsilon_{\rm def}$ for
every $k$, so the accumulated perturbation terms sum to
$\sum_kO(\epsilon_k\epsilon_{\rm def}+\epsilon_{\rm def}^2)
=O(\epsilon_{\rm def}^2)$ (constants depending on the fixed $K$) and
$\delta_K=\sum_kB_{K-1}\cdots B_{k+1}e_k+O(\epsilon_{\rm def}^2)$.  A final
expansion of the $C^2$ report at $x_K$,
$R(x_K+\delta_K,\xi)-R(x_K,\xi)=A_K\delta_K+O(\|\delta_K\|^2)$, together
with $A_{k+1}=A_KB_{K-1}\cdots B_{k+1}$ and
$\|\delta_K\|=O(\epsilon_{\rm def})$, proves \eqref{eq:adjointsum}.
\end{proof}

\begin{theorem}[First-order loss consistency]
\label{thm:loss}
Under the conditions of Theorem~\ref{thm:adjointsum} and
$\Lambda=\Lambda^\top\succ0$ bounded, for small $\epsilon_{\rm def}$
\begin{equation}
\left|
\norm{R(x_K^{\rm ex},\xi)-R(x_K,\xi)}_\Lambda^2
-
\big\|{\textstyle\sum_k}A_{k+1}e_k\big\|_\Lambda^2
\right|
\le C_{\Lambda,K}\epsilon_{\rm def}^3 .
  \label{eq:surrogateconsistency}
\end{equation}
\end{theorem}
\begin{proof}
Set $Z=\sum_kA_{k+1}e_k$ and let $q=\mathcal R$ be the remainder of
Theorem~\ref{thm:adjointsum}, so that
$R(x_K^{\rm ex},\xi)-R(x_K,\xi)=Z+q$ with
$\norm{Z}\le\sum_k\norm{A_{k+1}}\,\norm{e_k}=O(\epsilon_{\rm def})$ by the
bounded transition products and $\norm{q}\le C_K\epsilon_{\rm def}^2$.
Expanding,
$\norm{Z+q}_\Lambda^2-\norm{Z}_\Lambda^2
=2\inner{Z}{\Lambda q}+\norm{q}_\Lambda^2$, where
$|2\inner{Z}{\Lambda q}|\le2\norm{\Lambda}\norm{Z}\norm{q}
=O(\epsilon_{\rm def}^3)$ and
$\norm{q}_\Lambda^2=O(\epsilon_{\rm def}^4)\le O(\epsilon_{\rm def}^3)$ for
$\epsilon_{\rm def}\le1$.  The constants are uniform on the tube of
Assumption~\ref{ass:regularity}, so the bound survives taking expectations.
\end{proof}

The next result connects the exact and implemented sums, writing
$\widehat Z=\sum_kA_{k+1}\widehat e_k$ and $Z=\sum_kA_{k+1}e_k$.

\begin{proposition}[Richardson-substituted first-order term]
\label{prop:substitution}
Consider a refining family of admissible grids whose clock speeds satisfy
$0<\theta_{\min}\le\theta\le\theta_{\max}$, with transition products,
adjoint norms, and local-defect constants uniformly bounded over the
family.  Then, as $h=1/K\to0$,
$\|\widehat Z-Z\|=O(h^{r+1})$ while $\|Z\|=O(h^{r})$, and consequently,
whenever $\|Z\|=\Theta(h^{r})$,
\[
\Big|\;\big\|\widehat Z\big\|_\Lambda^2-\norm{Z}_\Lambda^2\;\Big|
=O(h^{2r+1})
=\norm{Z}_\Lambda^2\,O(h).
\]
\end{proposition}
\begin{proof}
By Proposition~\ref{prop:richardson},
$\|\widehat e_k-e_k\|\le C\Delta_k^{r+2}$ with $C$ uniform over the family.
Since $\Delta_k\le\theta_{\max}h$ and $\sum_k\Delta_k=T$, bounded adjoints
give
$\|\widehat Z-Z\|\le C'\sum_k\Delta_k^{r+2}
\le C'(\theta_{\max}h)^{r+1}\sum_k\Delta_k=O(h^{r+1})$, and likewise
$\|Z\|\le C''\sum_k\Delta_k^{r+1}=O(h^{r})$.  The loss comparison follows
from
$|\,\|\widehat Z\|_\Lambda^2-\|Z\|_\Lambda^2\,|
\le2\norm{\Lambda}\norm{Z}\|\widehat Z-Z\|
+\norm{\Lambda}\|\widehat Z-Z\|^2$, whose leading term is
$O(h^{r})O(h^{r+1})=O(h^{2r+1})$.
\end{proof}

Under stronger cancellation, $\|Z\|=o(h^r)$ and the two terms may be
comparable.  For a finite high-NFE reference, write $x^\star=x_K^{\rm ex}$
for the exact-flow endpoint and
$\eta_{\rm ref}=R(x^\star,\xi)-R(x^{\rm ref},\xi)$; the squared paired loss
is $\|Z\|_\Lambda^2-2\inner{Z}{\Lambda\eta_{\rm ref}}
+\|\eta_{\rm ref}\|_\Lambda^2
+O\!\big((\|Z\|+\|\eta_{\rm ref}\|)\epsilon_{\rm def}^2
+\epsilon_{\rm def}^4\big)$, so the quadratic
term leads when $\|Z\|=\Theta(\epsilon_{\rm def})$ and
$\|\eta_{\rm ref}\|=o(\epsilon_{\rm def})$; otherwise
rankings are finite-reference statements, and the doubled-resolution check
in Section~\ref{sec:experiments} preserves every displayed winner.

These local results justify the proposal surrogate, not global
optimality of the two-stage search.  With
$z_k=A_{k+1}e_k$,
\begin{equation}
  \big\|{\textstyle\sum_k}z_k\big\|_\Lambda^2
  =
  \underbrace{\sum_k\norm{z_k}_\Lambda^2}_{D}
  +\underbrace{2\sum_{i<j}\inner{z_i}{\Lambda z_j}}_{C}.
  \label{eq:cross}
\end{equation}
A diagonal approximation keeps only $D$; \art\ further drops the adjoint and
minimizes state-space defects.  The next result quantifies what this
discards when transported defects reinforce, the regime ($C>0$) we observe
throughout Section~\ref{sec:experiments}.

\begin{proposition}[Aligned defects: optimal grids and diagonal inefficiency]
\label{prop:aligned}
Fix $r\ge1$, $T>0$, and $a_k>0$.  Suppose transported defects follow the
frozen-coefficient power model with common direction:
$z_k(\Delta)=\Delta_k^{r+1}a_kv$, $\norm{v}_\Lambda=1$.  On
$\mathcal D_K(T)$:
\emph{(i)}
$\mathcal L_{\rm term}(\Delta)=\big(\sum_ka_k\Delta_k^{r+1}\big)^2$ is
uniquely minimized at
$\Delta_k^\star=T\,a_k^{-1/r}/\sum_ja_j^{-1/r}$, with first-order report
error $\sum_ka_k(\Delta_k^\star)^{r+1}=T^{r+1}\big(\sum_ka_k^{-1/r}\big)^{-r}$;
\emph{(ii)} the diagonal criterion
$\sum_ka_k^2\Delta_k^{2(r+1)}$ is uniquely minimized at
$\Delta_k^\diamond=T\,a_k^{-2/(2r+1)}/\sum_ja_j^{-2/(2r+1)}$;
\emph{(iii)} with $b_k=a_k^{-1/(2r+1)}$,
\begin{equation}
\rho
:=\frac{\sum_ka_k(\Delta_k^\diamond)^{r+1}}{\sum_ka_k(\Delta_k^\star)^{r+1}}
=\frac{\big(\sum_kb_k\big)\big(\sum_kb_k^{(2r+1)/r}\big)^{r}}
      {\big(\sum_kb_k^2\big)^{r+1}}
\;\ge\;1,
\label{eq:rho}
\end{equation}
with equality iff all $a_k$ are equal; $\rho$ is the first-order report-error
factor, and $\rho^2$ is the corresponding squared-error inflation.
\end{proposition}
\begin{proof}
Minimizing $\mathcal L_{\rm term}$ is equivalent to minimizing its positive
square root $\sum_ka_k\Delta_k^{r+1}$, which, like the diagonal criterion, is
strictly convex on the simplex closure; no boundary point is optimal,
since moving mass into an empty cell has zero marginal cost against a
strictly positive saving elsewhere.  Lagrange stationarity
gives $(r{+}1)a_k\Delta_k^r=\lambda$, hence $\Delta_k\propto a_k^{-1/r}$, and
$2(r{+}1)a_k^2\Delta_k^{2r+1}=\lambda'$, hence
$\Delta_k\propto a_k^{-2/(2r+1)}$; substitution gives the stated values and,
using $a_k=b_k^{-(2r+1)}$,
\begin{align*}
\sum_ka_k(\Delta_k^\diamond)^{r+1}
&=\frac{T^{r+1}\sum_kb_k}{\big(\sum_kb_k^2\big)^{r+1}},\\
\sum_ka_k(\Delta_k^\star)^{r+1}
&=\frac{T^{r+1}}{\big(\sum_kb_k^{(2r+1)/r}\big)^{r}},
\end{align*}
which yield \eqref{eq:rho}.  For the bound, write
$b_k^2=b_k^{1/(r+1)}\cdot b_k^{(2r+1)/(r+1)}$ and apply H\"older with
exponents $(r{+}1,\ (r{+}1)/r)$:
\[
\sum_kb_k^2\le\Big(\sum_kb_k\Big)^{1/(r+1)}
\Big(\sum_kb_k^{(2r+1)/r}\Big)^{r/(r+1)};
\]
raising to the $(r{+}1)$-th power gives $\rho\ge1$.  Equality requires
$b_k\propto b_k^{(2r+1)/r}$, i.e., $b_k$ constant.  For $r=1$ the bound is
Cauchy--Schwarz: $(\sum b^2)^2\le(\sum b)(\sum b^3)$.
\end{proof}

\begin{corollary}[Small-dispersion inefficiency]
\label{cor:dispersion}
Let $r=1$ and write $b_k=\bar b(1+x_k)$ with $\sum_kx_k=0$,
$m_2=K^{-1}\sum_kx_k^2$, and $m_3=K^{-1}\sum_k|x_k|^3$.  Then
$\rho=\big(\sum_kb_k\big)\big(\sum_kb_k^3\big)/\big(\sum_kb_k^2\big)^2
=1+m_2+O(m_3+m_2^2)$.
\end{corollary}
\begin{proof}
For $j\in\{1,2,3\}$,
$K^{-1}\sum_k(1+x_k)^j=1+\binom{j}{2}m_2+O(m_3)$ because $\sum_kx_k=0$;
the powers of $K$ and $\bar b$ cancel in $\rho$, so
$\rho=(1+3m_2)/(1+m_2)^2+O(m_3)=1+m_2+O(m_3+m_2^2)$.
\end{proof}

\begin{remark}
\label{rem:aligned}
\art\ additionally replaces the adjoint-weighted $a_k=\|A_{k+1}c_k\|$ by the
untransported $\|c_k\|_2$, compounding the exponent misallocation of
(iii) with a weight mismatch whenever transport gains vary over $k$.
Locally around homogeneous coefficients, $\rho-1$ grows to leading order
with the dispersion of $b_k=a_k^{-1/3}$ (Corollary~\ref{cor:dispersion}),
so the model predicts larger schedule-choice gains where adjoint-weighted
defects are more heterogeneous.  Estimating
$a_k=\|A_{k+1}\widehat e_k\|/\Delta_k^{2}$ on the UNIF grid at $K=8$ gives
post-hoc evaluation-panel estimates
$\rho=1.026$ for Regime-SV and $\rho=1.010$ for Mag7---squared-error
inflations of $5.3\%$ and $2.0\%$ from step misallocation alone,
qualitatively paralleling the larger Regime-SV gains in
Table~\ref{tab:finance-errors}---with mean alignment cosines $0.84$ and
$0.95$ supporting the aligned idealization.  Alignment is an idealization of the reinforcing regime; the proposition
concerns first-order surrogates, not the full pipeline.
\end{remark}

\section{Numerical Evaluation}
\label{sec:experiments}
We ask three questions.  Does two-stage \tapes\ reduce paired report error
at fixed deployment NFE?  Do paired uncertainty, reference-resolution
checks, and matched construction diagnostics support that result?  And do
report-space gains carry financial meaning
(Section~\ref{sec:downstream})?

\subsection{Data and Protocol}
All deployments use normalized progress $T=1$, Euler grids with
$K\in\{4,8,12,16,24,32\}$, and a paired UNIF $K_{\rm ref}=1024$ numerical
reference; Richardson probes use one full versus two half steps with $r=1$.
The same initial noise and, when applicable, context are used for every
grid; construction uses development items only.  Regime-SV evaluation uses
376 independent noise paths; Mag7 uses 120 chronologically later test
contexts.

\paragraph{Regime-SV}
With simulation seed 202 we generate 2{,}600 five-asset observations from
two regimes with $P_{00}=P_{11}=.975$: variance-standardized Student-$t_5$
innovations; a seeded rank-three factor covariance shrunk as
$0.35\Sigma+0.65I$ in the calm regime and $0.75\bm1\bm1^\top+0.25I$ under
stress; base volatility $0.011$ with stress target $2.8\times$; log
volatility with persistence $0.96$, target weight $0.04$, absolute-return
feedback $0.08$, leverage $-0.10$, innovation s.d.\ $0.035$.
Training-prefix standardization ends at observation 1819; length-eight
windows with stride four and an eight-step embargo yield 452/92/94
train/validation/test windows, and four paired repeats give 376 evaluation
noises.  The AdamW-trained score model (seed 5428) has two layers, width
192, time embedding 64, batch size 128, 60 epochs, learning rate
$3\times10^{-4}$, VP parameters $\beta_{\min}=.05$, $\beta_{\max}=3.0$,
score clipping 8.0.

\paragraph{Mag7.}
We use complete-case, auto-adjusted Yahoo Finance closes for seven stocks:
AAPL, MSFT, NVDA, AMZN, GOOGL, META, and TSLA.  Closes queried over May 18,
2012--May 25, 2026 yield 3{,}522 log returns dated May 21, 2012--May 22,
2026; these are standardized on the training prefix.  The conditional
model (seed 5431) maps 32 past returns to eight future returns: width 384,
three layers, time embedding 64, batch size 128, 120 epochs, learning rate
$2.5\times10^{-4}$, $\beta_{\min}=.02$, $\beta_{\max}=1.75$, score clipping
2.0.  Stride-four windows with an eight-step embargo yield 605/118/120
train/validation/test items; the future targets of the 120 test items span June 14, 2024
through May 20, 2026.

Test uncertainty uses 500 paired bootstrap resamples per contrast: iid for
Regime-SV and non-circular moving blocks of four consecutive Mag7 contexts
\cite{kunsch1989jackknife}.

\subsection{Paired Report Accuracy}
\begin{table}[t]
\centering
\caption{Evaluation-panel means of squared residual norms in normalized
report coordinates against a paired UNIF $K_{\rm ref}=1024$ reference.
Des.\ and Aud.\ denote $E_{\rm design}$ and $E_{\rm audit}$; lower is
better; bold marks the smallest displayed mean.  The five schedule
implementations are defined in Section~\ref{sec:baselines}; Mag7 contexts are
chronologically held out.}
\label{tab:finance-errors}
\begin{tabular}{@{}llccccc@{}}
\toprule
$K$ & & \tapes & \art & \ays & \edm & UNIF\\
\midrule
\multicolumn{7}{@{}l}{\textit{Regime-SV}}\\
4  & Des. & \textbf{2.63630} & 2.87576 & 2.82134 & 2.88594 & 3.06231\\
   & Aud. & \textbf{1.37777} & 1.48623 & 1.45675 & 1.48815 & 1.57483\\
8  & Des. & \textbf{0.51195} & 0.54412 & 0.56910 & 0.56974 & 0.55464\\
   & Aud. & \textbf{0.28801} & 0.30277 & 0.31217 & 0.31568 & 0.31326\\
12 & Des. & \textbf{0.18980} & 0.22337 & 0.21742 & 0.22781 & 0.21742\\
   & Aud. & \textbf{0.11503} & 0.12706 & 0.12824 & 0.13060 & 0.12824\\
16 & Des. & \textbf{0.09614} & 0.12196 & 0.11583 & 0.12319 & 0.11583\\
   & Aud. & \textbf{0.06049} & 0.06956 & 0.06900 & 0.07096 & 0.06900\\
24 & Des. & \textbf{0.03626} & 0.05183 & 0.04800 & 0.05098 & 0.04800\\
   & Aud. & \textbf{0.02513} & 0.02980 & 0.02904 & 0.03006 & 0.02904\\
32 & Des. & \textbf{0.01964} & 0.02832 & 0.02585 & 0.02762 & 0.02585\\
   & Aud. & \textbf{0.01369} & 0.01637 & 0.01580 & 0.01642 & 0.01580\\
\midrule
\multicolumn{7}{@{}l}{\textit{Mag7}}\\
4  & Des. & \textbf{0.69841} & 0.70536 & 0.71163 & 0.70898 & 0.72731\\
   & Aud. & \textbf{0.73308} & 0.74267 & 0.73950 & 0.73954 & 0.76817\\
8  & Des. & \textbf{0.17408} & 0.17501 & 0.17854 & 0.17696 & 0.17889\\
   & Aud. & \textbf{0.19180} & 0.19398 & 0.19370 & 0.19365 & 0.20127\\
12 & Des. & \textbf{0.07681} & 0.07706 & 0.07902 & 0.07819 & 0.07863\\
   & Aud. & \textbf{0.08629} & 0.08701 & 0.08720 & 0.08705 & 0.09043\\
16 & Des. & \textbf{0.04295} & 0.04310 & 0.04418 & 0.04370 & 0.04376\\
   & Aud. & \textbf{0.04862} & 0.04907 & 0.04915 & 0.04908 & 0.05097\\
24 & Des. & \textbf{0.01883} & 0.01885 & 0.01937 & 0.01914 & 0.01913\\
   & Aud. & \textbf{0.02148} & 0.02169 & 0.02173 & 0.02170 & 0.02252\\
32 & Des. & \textbf{0.01042} & 0.01044 & 0.01074 & 0.01060 & 0.01059\\
   & Aud. & \textbf{0.01198} & 0.01207 & 0.01210 & 0.01208 & 0.01254\\
\bottomrule
\end{tabular}
\end{table}

\begin{figure}[t]
\centering
\includegraphics[width=.88\columnwidth]{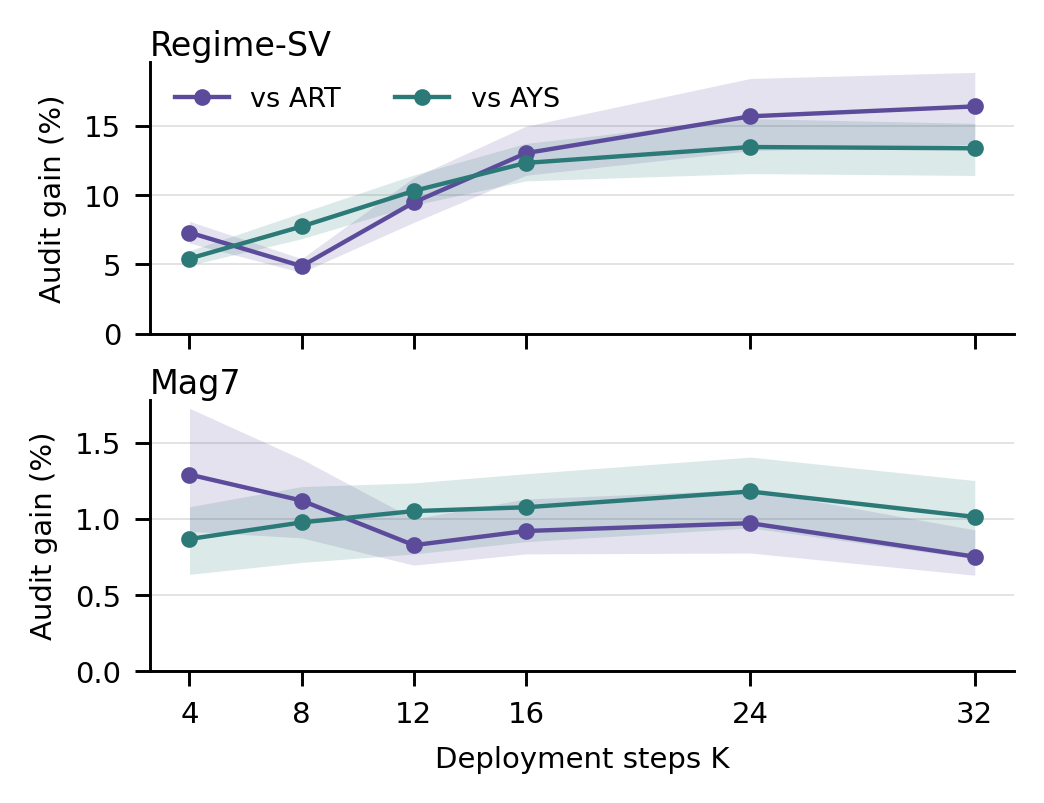}
\caption{Paired audit-error reductions across deployment budgets; positive
values favor \tapes\ over the \art\ and \ays\ baselines.  Shading gives
pointwise 95\% paired-bootstrap intervals on the
Table~\ref{tab:finance-errors} panel (iid for Regime-SV; non-circular
blocks of four Mag7 contexts).  All 24 audit intervals exclude zero;
because budgets reuse items, the intervals are not simultaneous.}
\Description{Two line plots show positive TAPES audit-error gains against
the ART and AYS baselines for every deployment budget.  Regime-SV gains
range from about 5 to 17 percent and Mag7 gains from about 0.75 to 1.3
percent; every pointwise confidence band remains above zero.}
\label{fig:audit-gains}
\end{figure}

Within the reported five-pipeline panel, the full \tapes\ pipeline has the
lowest sample mean in all 24 dataset--$K$--report rows of
Table~\ref{tab:finance-errors}.  Paired
pointwise 95\% intervals for \tapes-minus-comparator differences lie below
zero in all 24 comparisons with \ays\ and 23 of 24 with \art\ (exception:
Mag7 $K=24$ design), including all 12 audit rows for both comparators.
Figure~\ref{fig:audit-gains} shows the paired audit comparisons; because
budgets and reports share items, the 24-row consistency count is
descriptive.  For a familywise summary, define
$G=100\,\operatorname{mean}_{K}(1-\bar E_{\mathrm{TAPES},K}/\bar E_{B,K})$
and jointly resample the same item indices at every $K$ from the direct
$K_{\rm ref}=2048$ reevaluation (100{,}000 resamples).
Table~\ref{tab:aggregate-gain} reports percentile intervals at 98.75\%
marginal coverage, controlling the four-contrast familywise error at 5\% by
Bonferroni; every interval is above zero.

\begin{table}[t]
\centering
\caption{Cross-budget relative audit-error reduction at $K_{\rm ref}=2048$.
Intervals have 98.75\% marginal coverage for a four-contrast Bonferroni
familywise check.}
\label{tab:aggregate-gain}
\begin{tabular}{@{}llrr@{}}
\toprule
Data & Baseline & $G$ (\%) & 98.75\% interval\\
\midrule
Regime-SV & \art & 11.04 & $[9.88,12.15]$\\
Regime-SV & \ays & 10.37 & $[9.58,11.12]$\\
Mag7 & \art & 0.975 & $[0.753,1.219]$\\
Mag7 & \ays & 1.016 & $[0.749,1.233]$\\
\bottomrule
\end{tabular}
\end{table}

\paragraph{Direct doubled-reference check.}
Reevaluating every saved grid on newly generated paired UNIF
$K_{\rm ref}=2048$ paths (all 376 Regime-SV and 120 Mag7 items) preserves
every displayed winner and every non-tied ordering in all 24 rows; 47 of 48 pointwise
comparator intervals stay below zero (10{,}000 resamples; the exception
remains Mag7 $K=24$ design versus \art), and Mag7 block lengths 2, 4, and 8
each leave all 12 audit contrasts below zero.  This stabilizes the finite-reference
conclusion.

\paragraph{Beyond Euler: a Heun probe.}
To probe transfer beyond Euler, we reran the construction pipeline on
Regime-SV with Heun deployment ($r=2$, NFE $=2K$) at $K\in\{4,8\}$: 188
paired evaluation noises, a fresh UNIF $K_{\rm ref}=1024$ Heun reference, 32
development items, and a reduced candidate library.  Across the same five
reported pipelines, \tapes\ has the lowest audit mean at both budgets
($.01226$ versus $.01252$ for the next-lowest reported pipeline at $K=4$;
$.00079$ versus $.00083$ at $K=8$), with paired 95\% audit intervals below
zero against the next-lowest reported pipeline and \art.  This is a
pipeline-level probe, not an exhaustive comparison of internal candidates.

\subsection{What Each Construction Stage Contributes}
\label{sec:matched}
Although the generator is frozen, its low-NFE integration grid is not part
of the checkpoint.  For each $K$, full \tapes\ constructs that numerical
control offline: it fits four base clocks to the positive cosine
representation, takes one Richardson/terminal-adjoint proposal on 64
development items using the design report, applies at most eight Regime-SV
or four Mag7 L-BFGS-B iterations to paired audit-report error, and selects
the lowest-development-error candidate.  That final grid---the \tapes\ row
in Tables~\ref{tab:finance-errors} and~\ref{tab:downstream}---is then frozen;
no generator training occurs, and deployment still uses exactly $K$ model
calls.

Table~\ref{tab:attribution} replays the implemented construction stages without
retraining and compares full \tapes\ with two ablations.  \emph{Before
direct refinement} is the exact post-adjoint ancestor of the selected final
grid, so it tests whether the last direct-refinement stage helps that
selected path.  \emph{Without adjoint proposal} starts from the same fitted
base and uses the same clock basis, floor, 64-item panel, paired reference,
audit objective, and refinement cap, but skips the adjoint update.  The
resulting grids are evaluated on the same 376 Regime-SV and 120
chronological Mag7 items as Table~\ref{tab:finance-errors}.

\begin{table}[b]
\centering
\caption{Matched construction-stage ablations.  Cells are mean paired
evaluation error of full \tapes\ minus the named variant; negative favors
full \tapes.  Des./Aud. denote design/audit errors against paired UNIF
$K_{\rm ref}=1024$ references.}
\label{tab:attribution}
\begin{tabular}{@{}lrrrr@{}}
\toprule
& \multicolumn{2}{c}{\shortstack{Before direct\\refinement}}
& \multicolumn{2}{c}{\shortstack{Without adjoint\\proposal}}\\
\cmidrule(lr){2-3}\cmidrule(l){4-5}
$K$ & Des. & Aud. & Des. & Aud.\\
\midrule
\multicolumn{5}{@{}l}{\textit{Regime-SV}}\\
4  & $-.003556$ & $-.001647$ & $-.064874$ & $-.025345$\\
8  & $+.001815$ & $+.000140$ & $-.008672$ & $-.003370$\\
12 & $-.020906$ & $-.006569$ & $-.034928$ & $-.014052$\\
16 & $-.000004$ & $-.000001$ & $-.009513$ & $-.002687$\\
24 & $+.000453$ & $-.000099$ & $-.004510$ & $-.001061$\\
32 & $-.001243$ & $-.000034$ & $-.002219$ & $-.000454$\\
\midrule
\multicolumn{5}{@{}l}{\textit{Mag7}}\\
4  & $-.000425$ & $-.001864$ & $+.003008$ & $+.000578$\\
8  & $-.003105$ & $-.000973$ & $-.004460$ & $-.001892$\\
12 & $-1.0\!\times\!10^{-8}$ & $-1.4\!\times\!10^{-8}$ & $-.000411$ & $-.000376$\\
16 & $-.000015$ & $-.000049$ & $-.000116$ & $-.000018$\\
24 & $+.000018$ & $-.000044$ & $-.000005$ & $-.000086$\\
32 & $-4.5\!\times\!10^{-8}$ & $-2.8\!\times\!10^{-8}$ & $-.000053$ & $-.000048$\\
\bottomrule
\end{tabular}
\end{table}

Full \tapes, not either ablation, is the deployed method.  In 10{,}000 shared
resamples, full-minus-without-adjoint is negative in 22/24 means (20
intervals below zero, three containing zero); the sole interval above zero is
Mag7 $K=4$ design ($+.003008$, 95\% CI $[+.002351,+.003773]$).
Full-minus-before-refinement is negative in 20/24 means (17 intervals below,
four containing zero, three above).  Under the implemented caps, the adjoint
proposal therefore helps more consistently, while direct refinement helps
most but not every row.
Reevaluation at $K_{\rm ref}=2048$ preserves all 48 signs and interval
classes.  The matched without-adjoint comparison is the closest controlled
diagnostic of the proposal, but the full route searches additional proposals
and inherited base tuning is uncharged; neither ablation is equal-compute or causal.

\subsection{Why Terminal Transport Matters: Mechanism}
\label{sec:mechanism}
\begin{figure}[t]
\centering
\includegraphics[width=\columnwidth]{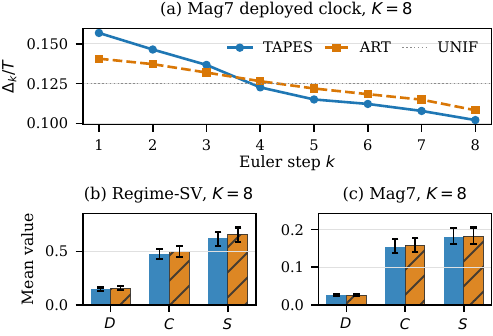}
\caption{Final-grid mechanism at $K=8$.  (a)~Deployed increments
$\Delta_k/T$ for Mag7; UNIF dotted.  (b,~c)~Evaluation-panel means and
pointwise 95\% intervals for the diagonal term $D$, signed cross term $C$,
and $S=D+C$ in normalized design-report coordinates.  $C>0$ indicates
cross-step reinforcement.}
\Description{Panel a shows the TAPES clock on Mag7 taking larger early
steps and smaller late steps than the ART clock, crossing it near step
four.  Panels b and c show grouped bars where the cross term C is three to
six times larger than the diagonal term D for both datasets, with TAPES
bars slightly below ART bars for D, C, and S.}
\label{fig:mechanism}
\end{figure}

For $z_k=A_{k+1}\widehat e_k$ in normalized design-report coordinates,
decompose $S=\|\sum_kz_k\|^2=D+C$ as in \eqref{eq:cross}.
Figure~\ref{fig:mechanism} shows the deployed $K=8$ Mag7 clock and the
paired decomposition.  The selected \tapes\ clock crosses
the \art\ clock, moving faster through early diffusion time and
allocating more resolution to later intervals, consistent with the
report-aware transported-defect objective.  Cross terms dominate and are
positive---$\bar C/\bar D=3.27$
(Regime-SV, $K=8$), $6.15$ (Mag7, $K=8$), $27.2$ (Mag7, $K=32$)---so the
observed regime is \emph{reinforcement}: a diagonal criterion ignores
most of the objective.
\tapes\ reduces $D$, $C$, and $S$ relative to \art\ in all three displayed settings;
the paired 95\% intervals for the $S$ difference are
$[-.04607,-.03288]$ (Regime-SV) and $[-.00244,-.00138]$ (Mag7) at $K=8$.
A direct theory-to-error audit at $K_{\rm ref}=2048$ supports the
first-order sum on the frozen final grids: all 1{,}488 item--grid inner
products between $\widehat Z=\sum_kA_{k+1}\widehat e_k$ and the realized
report difference are positive, with mean cosines $.998$/$.9997$ and
median relative residuals $.124$/$.025$.  These are surrogate-geometry diagnostics on final grids, not
attributions to one construction stage.

\subsection{Exploratory 21-Day Numerical and Allocation Sensitivity}
\label{sec:downstream}
Does report-space fidelity carry through a decision layer?
On a separately frozen Mag7 checkpoint under the fixed
21-day-horizon ($H=21$) construction protocol, we evaluate whether the
numerical ordering extends: across 52 successive 21-trading-day blocks of 1{,}024 paired paths
with a fresh UNIF $K_{\rm ref}=1024$ reference, the $H=21$ \tapes\ grid has
the lowest displayed $K=8$ design and audit errors (Table~\ref{tab:downstream}),
and all four moving-block 95\% intervals for \tapes\ minus \art\ or \ays{}
lie below zero (10{,}000 resamples; largest upper endpoint $-.00314$).
The panel overlaps one development context and six candidate-scoring
contexts, so this is a numerical sensitivity analysis rather than an
independent holdout.

The same five grids then drive diffusion-model-assisted scenario training of
a Black--Scholes/no-factor continuous-time mean--variance (CTMV)
policy-gradient allocator using the Markowitz objective
\cite{markowitz1952portfolio}.  The downstream allocation experiment follows the
continuous-time mean--variance reinforcement-learning methodology developed in
\cite{huang2022meanvariance,huang2024meanvariance}, with identical allocation
settings held fixed across sampling grids.  At each of 52 consecutive 21-trading-day
rebalances (January 3, 2022--April 13, 2026), the frozen conditional generator
maps the preceding 32-day context and shared Gaussian noises to $B=1{,}024$
21-day scenarios.  Each method's actor--critic is warm-started across blocks
and fine-tuned for 1{,}000 policy-gradient updates (batch 128) toward
$z_H=1.02844$ (40.0\% annualized), after the same fixed tail-stress input
transform.  Its greedy policy yields fully invested long--short weights
(sum one; no exposure bound), held over the next 21 realized days and scored
at zero transaction cost.  This is scenario-trained RL, not a closed-form
sample mean--covariance Markowitz solve.  Because the allocation configuration
was selected on this same 52-block period, its metrics are conditional,
post-selected descriptions, not out-of-sample evidence.  Only the grid and
hence its scenario panel vary.  The final realized block ends May 11, 2026.
The $H=21$ checkpoint's training targets end by December 31, 2021, and its
return standardization uses data only through that date, before the first
allocation block on January 3, 2022.
Table~\ref{tab:downstream} places the resulting allocation metrics beside
the bridge errors.  Within the displayed five-pipeline panel, the Sharpe and Sortino ordering matches
the report-error ordering (\tapes\ $>$ \art\ $>$
\ays\ $>$ \edm\ $>$ UNIF).  Among the five pipelines, \tapes\ also achieved
the lowest maximum drawdown and the shortest recovery time.

\begin{table}[t]
\centering
\caption{$H=21$ numerical bridge and exploratory selected-setting allocation sensitivity at
$K=8$.  Des./Aud.: mean report errors against paired UNIF
$K_{\rm ref}=1024$ references (52 successive 21-day blocks $\times$ 1{,}024 paths).
Ret.\ is annualized return in percent; Sort.\ and Cal.\ are Sortino and
Calmar ratios, realized by the identical scenario-trained CTMV
policy-gradient allocator over the same 52 blocks.  Bold marks the best
displayed value.}
\label{tab:downstream}
\begin{tabular}{@{}lcccccc@{}}
\toprule
Grid & Des. & Aud. & Ret.\ (\%) & Sharpe & Sort. & Cal.\\
\midrule
\tapes & \textbf{.24532} & \textbf{.27465} & \textbf{44.18} & \textbf{1.108} & \textbf{1.651} & \textbf{1.110}\\
\art   & .24875 & .27826 & 40.83 & 1.043 & 1.556 & 1.026\\
\ays   & .24941 & .27911 & 39.87 & 1.023 & 1.530 & 1.002\\
\edm   & .25323 & .28178 & 41.20 & 1.019 & 1.507 & 1.023\\
UNIF   & .26096 & .29257 & 40.42 & 0.986 & 1.456 & 0.991\\
\bottomrule
\end{tabular}
\end{table}

\begin{figure}[t]
\centering
\includegraphics[width=.92\columnwidth]{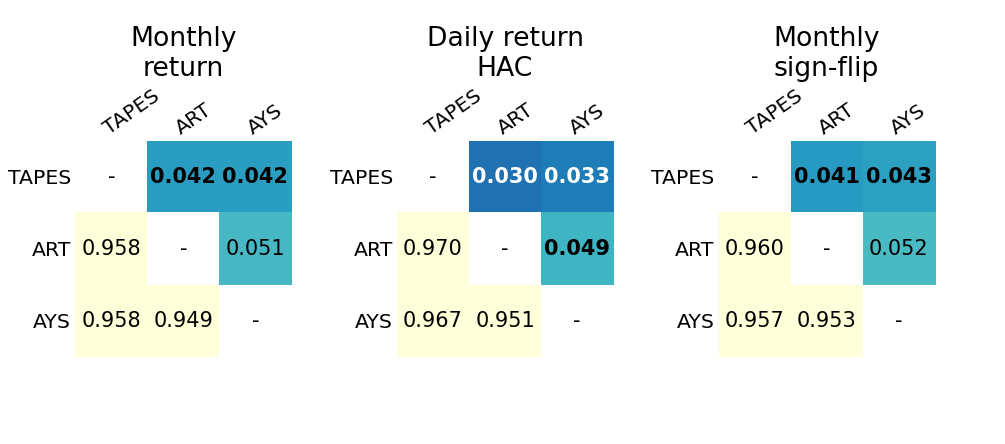}
\caption{Conditional one-sided pairwise $p$-values on the selected panel
under 21-day paired-block, daily Newey--West/HAC, and block sign-flip tests.
Entry $(i,j)$ tests row~$i$ over column~$j$; bold denotes $p<.05$.
All \tapes\ entries against \art\ and \ays\ are below $.05$, but these
unadjusted post-selection results are descriptive.}
\Description{Three heatmaps of one-sided pairwise p-values for TAPES, ART,
and AYS under monthly paired, daily HAC, and sign-flip tests.  The TAPES
row is below 0.05 against ART and AYS in all three panels, with values
0.042 and 0.042, 0.030 and 0.033, and 0.041 and 0.043.}
\label{fig:downstream-tests}
\end{figure}

Conditional on the selected 52-block panel, Figure~\ref{fig:downstream-tests}
reports unadjusted one-sided $p$-values under paired 21-day block returns,
daily Newey--West/HAC returns \cite{newey1987hac}, and block sign flips; all
six displayed entries comparing \tapes\ with \art\ or \ays\ are below $.05$.  Charging
5--20 bps per unit of one-way turnover preserves the displayed \tapes-first
block-return Sharpe ordering.  These conditional diagnostics neither undo
setting selection nor establish a trading advantage.

\section{Limitations and Conclusion}
\paragraph{Limitations.}
The evidence covers two data sources, three frozen checkpoints, Euler plus a
small Heun probe, and one $H=21$ extension.  The high-NFE reference is a
numerical target from the same generator, not the market law; intervals
condition on frozen checkpoints and selected grids.  Some methods use
audit-family data during development, and their stated procedures do not
equalize offline compute.  Broader comparison across alternative solver and
schedule stacks, a causal marginal adjoint effect, and replication across
training runs remain open.  The allocation configuration
was selected on the same return period; its diagnostics are exploratory, not
out-of-sample evidence.  Other assets, periods, targets, costs, and
constraints remain untested.

\paragraph{Conclusion.}
Within those boundaries the evidence is consistent at every layer.  Theory
identifies the transported signed sum as the first-order terminal-report
error; the aligned model gives an inefficiency factor qualitatively consistent
with the larger Regime-SV gains, plus a transport mismatch for the
untransported defect criterion.  Within the reported five-pipeline panel, \tapes\ has the lowest
mean paired error in all 24 rows at fixed deployment NFE, with all four
cross-budget Bonferroni intervals
excluding zero and every displayed ordering stable under doubled resolution.
A separate $H=21$ sensitivity preserves this numerical ordering; one
selected-setting allocation exercise illustrates---without confirmatory
performance claims---how grid differences can propagate through a fixed
decision pipeline.  \tapes\ changes no trained generator weights, solver, or deployment budget:
it reallocates numerical effort where the terminal report is most sensitive.

\begin{acks}
Yilie Huang acknowledges financial support from the Start-up Fund of
The Hong Kong Polytechnic University (Project ID: P0063874).
\end{acks}

\interlinepenalty=10000
\bibliographystyle{ACM-Reference-Format}
\bibliography{references}

\end{document}